\documentclass[aps,prd,showpacs,twocolumn,amssymb,superscriptaddress,floatfix]{revtex4}

\usepackage{graphicx}
\usepackage{longtable}

\usepackage{amsmath, amsthm, amssymb}
\usepackage{color}

\newcommand{\nl}{\nonumber\\ }
\newcommand{\pd}{\partial}

\def\be{\begin{eqnarray}}
\def\ee{\end{eqnarray}}

\def\lsim{\stackrel{\scriptstyle <}{\phantom{}_{\sim}}}
\def\gsim{\stackrel{\scriptstyle >}{\phantom{}_{\sim}}}

\begin{document}
\title{Shear and bulk viscosities for pure glue matter}

\author{A.S. Khvorostukhin}
\affiliation{GSI, Helmholtzzentrum f\"ur Schwerionenforschung GmbH,
Planckstrasse 1, 64291 Darmstadt, Germany}
\affiliation{ Joint Institute for Nuclear Research,  141980 Dubna, Russia}
\affiliation{Institute of Applied Physics, Moldova Academy of Science,
MD-2028 Kishineu, Moldova}

\author{V.D. Toneev} \affiliation{GSI, Helmholtzzentrum f\"ur
Schwerionenforschung GmbH, Planckstrasse 1, 64291 Darmstadt, Germany}\affiliation{ Joint Institute for Nuclear
Research,  141980 Dubna, Russia}

\author{D.N. Voskresensky} \affiliation{GSI, Helmholtzzentrum f\"ur
Schwerionenforschung GmbH, Planckstrasse 1, 64291 Darmstadt, Germany}
\affiliation{National Research Nuclear University "MEPhI",
Kashirskoe sh. 31, Moscow 115409, Russia}

\begin{abstract}
Shear $\eta$ and bulk $\zeta$ viscosities are calculated in a
quasiparticle model within a relaxation-time approximation  for
pure gluon matter. Below $T_c$, the confined sector is described
within a quasiparticle glueball model.
The constructed equation of state reproduces the first-order phase
transition for the glue matter. It is shown that with this
equation of state, it is possible to describe the temperature
dependence of the shear viscosity to entropy ratio $\eta/s$ and
the bulk viscosity to entropy ratio $\zeta/s$ in reasonable
agreement with available lattice data, but absolute values of the
$\zeta/s$ ratio underestimate  the upper limits of this ratio in
the lattice measurements typically by an order of magnitude.
\end{abstract}

\maketitle

\section{Introduction}

 The high-energy heavy-ion collisions at Super Proton Synchrotron (SPS) and Relativistic
Heavy-Ion Collider (RHIC) energies have shown evidence of a new
state of matter characterized by very low shear viscosity to
entropy density ratio $\eta/s$ similar to a nearly ideal
fluid~\cite{Sh05,GMcL05,Hei05,PC05}. Lattice calculations indicate
that the crossover region between hadron and quark-gluon matter
has been reached in these experiments. On the other hand, lattice
calculations performed in gluodynamics (GD) clearly demonstrate
the occurrence of there occurs the first-order phase transition.

The shear $\eta$ and bulk $\zeta$ viscosities are parameters that
quantify dissipative processes in the hydrodynamic evolution of a
fluid. It is known that the behavior of transport coefficients is
sensitive to the presence of phase transitions in a medium (see
\cite{Ka08,CKMcL06,ZH95,Lea07,ML37,PP06,KT07} and references
therein). Values of the bulk and shear viscosities near the
phase-transition critical temperature $T_c$ affect the
hydrodynamic evolution of the  medium and may influence
observables.

Lattice quantum chromodynamics (QCD) is the most powerful
technique to extract nonperturbative information on an equation of
state (EoS). When experimental data are lacking, lattice data are
often used to fit model parameters. For pure gluon SU(3) theory,
 the EoS was computed on the lattice
more than a decade ago~\cite{BEG95}. Recently,  much more accurate
data have been obtained~\cite{Pa09}.

Among  various existing phenomenological approaches, quasiparticle
(QP) models are used to reproduce results obtained in the lattice
QCD. In the case of GD, the QP models  rely on the assumption that
for the temperature $T$ above the critical one, $T>T_c$, the
system consists of a gas of massive deconfined  gluons.  In the
confined phase, at $T<T_c$,  the glue matter is considered as a
gas of massive glueballs.

In this paper we aim to investigate the behavior of viscosity
coefficients for a gluon system that exhibits a deconfinement
phase transition.  The phenomenological QP model is applied  to
describe available lattice  data on the EoS. Shear and bulk
viscosities are calculated within a relaxation-time approximation.

\section{Equation of state of glue matter }

 In the QP approach
the system of interacting gluons is treated as a gas of
noninteracting quasiparticles  with an effective mass $m_g(T)$,
which depends on $T$ as \cite{PC05}
 \be \label{m_g}
   m_g^2(T)=\frac{N_c}{6} \ g^2(T) \ T^2
 \ee
 with the temperature-dependent strong interaction constant
 \be \label{coupl}
 g^2(T)=\frac{48\pi^2}{11N_c \ \ln{[\lambda(T-T_s)/T_c)]^2}}~,
 \ee
where parameters $T_s/T_c=$0.5853,  $\lambda=$3.3 are taken to fit
the new lattice data, see below, and a
number of colors $N_c=$3. The energy
density and the pressure acquire then the following forms:
 \be
 \varepsilon_g(T) &=\frac{d_g}{
2\pi^2} \int_0^\infty p^2dp \frac{E}{ \exp(E/T) - 1} + B(T)\nl&
\equiv \varepsilon^{id}_g(T,m_g(T)) + B(T) , \label{Egl}
 \ee

 \be P_g(T) &=\frac{ d_g}{ 6\pi^2}\int_0^\infty p^2dp
\frac{p^2}{E}\frac{1}{\exp(E/T) - 1} - B(T)\nl & \equiv
P^{id}_g(T,m_g(T)) - B(T),
 \label{Pgl}
 \ee
 where the degeneracy factor $d_g = 2(N^2_c - 1)=$
16 for the $SU(3)$ gluodynamics, $\varepsilon^{id}_g$ and
$P^{id}_g$ are  the energy density and the pressure of the ideal gas
of massive gluons. The temperature-dependent function $B(T)$ in
Eq. (\ref{Egl}) results from the thermodynamical identity, see
Ref.~\cite{GY95},
 \be \label{b1}
  T \frac{dP}{dT} - P(T) = \varepsilon(T),
 \ee
  which leads to the equation for $B(T)$:
 \be \label{b2}
 \frac{dB(T)}{dT} = - \frac{\varepsilon^{id}_g -3P^{id}_g}{ m_g}
 \frac{dm_g}{dT}.
 \ee

Dealing only with gluon degrees of freedom, one assumes that the
matter at $T<T_c$ (the "hadronic" phase) consists of glueballs.
While the meson scattering amplitude is parametrically suppressed
as $1/N_c$, the scattering amplitude between glueballs scales as
$1/N_c^2$~\cite{Wi79} and therefore the system can be considered
as a noninteracting Bose gas of glueballs. Expected glueball
masses are high, $m_{gb}\gsim $1 GeV, and thereby only the
lowest-lying glueball states  contribute to the EoS at the
temperatures of our interest.
 It is difficult to single out which states of the observed
hadronic spectrum are glueballs because of a lack of  knowledge of
decay properties and the existence of a strong  mixing between
glueballs and quark states~\cite{MKV09}.  However,  using typical
constant values for the lowest-lying glueball masses within a
statistical model, one fails  to reproduce the strong increase of
thermodynamical variables near $T_c$~\cite{Bu09}. The $T$-behavior
of masses for two lowest-lying scalar $0^{++}$ and tensor $2^{++}$
glueballs was investigated on the lattice in~\cite{ISM02}.
Therefore, below we follow the SU(3) lattice GD results. It was
shown that the pole mass $m_{gb}(T)$, the Breit-Wigner mass
$\tilde{m}_{gb}(T)$, and the thermal width $\Gamma_{gb}$ are
linked as follows~:
 \be \label{Mglb}
m_{gb}(T)\approx
\tilde{m}_{gb}(T)-2T+\sqrt{4T^2-\Gamma_{gb}^2(T)}~.
 \ee
With the help of the {\it Ansatz} $\tilde{m}_{gb}(T)=m_{gb}^0$,
i.e., that the glueball Breit-Wigner  masses are given by the
particle data group (PDG) values,
 \be \label {Ggb}
  \Gamma_{gb}=b_{gb}
 (T-T_{gb})\
 \Theta (T-T_{gb}) \ \ {\rm for} \ \ T_{gb}<T<T_c
 \ee
 and recommended parameters $b_{gb}(0^{++})=$4.23 and
$b_{gb}(2^{++})=$7.152,  the relation (\ref{Mglb}) reproduces
 quite well the lattice results in the measured range
 $0.5T_c<T<T_c=265$ MeV~\cite{ISM02}. In our consideration,
 we limit ourselves to the two above-mentioned species of glueballs,
 the only ones for which lattice data are available.

With the temperature-dependent glueball masses,  a statistical
treatment of glueballs needs an additional requirement of
thermodynamic consistency. It has been satisfied in the same way
as outlined above for gluons; see Eqs. (\ref{b1}) and (\ref{b2})
above.

To describe glue matter in the whole range of temperatures,  we
use the first-order phase-transition model in accordance with
lattice results for GD. Thus,  one should conjugate the pure gluon
($g$) and the glueball ($gb$) phases by making use of the Gibbs
conditions at the transition:
 \be \label{Gib}
 T_c^{g}=T_c^{gb}\equiv T_c~, \ \ \ \ P_{g}(T_c)=P_{gb}(T_c)~.
 \ee
We use the value $T_c=$265 MeV for the first-order phase
transition, in agreement with the lattice SU(3)
GD~\cite{BEG95,Pa09,Mi07}.
\begin{figure}[thb]
\includegraphics[height=5.truecm,clip]{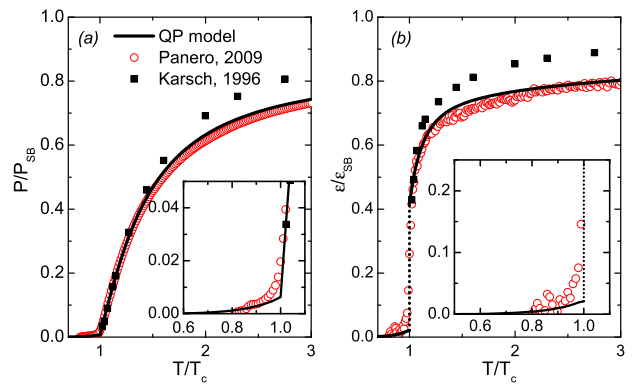}
\caption{ The reduced pressure (a) and the energy density (b) of
the glue  matter. The solid  line corresponds to taking into
account two glueballs, scalar $0^{++}$ with $m_{gb}^0=1470$ MeV
and tensor $2^{++}$ with $m_{gb}^0=2150$  MeV. Experimental points
are the old Karsch's (filled squares)~\cite{Ka89} and the new
Panero's (circles)~\cite{Pa09}
 lattice results. The region near $T_c$ is enlarged in the inset.
 }
 \label{pe}
\end{figure}

\begin{figure}[thb]
\includegraphics[height=5.truecm,clip]{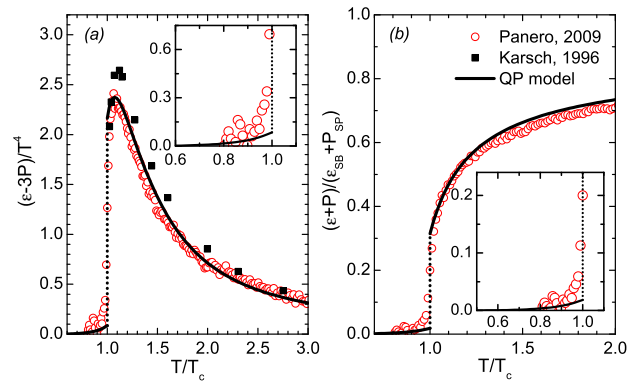}
\caption{
 The trace anomaly (a) and the reduced enthalpy or entropy
(b) of the glue matter. All notations are the same as in Fig.
\ref{pe}.
 }
 \label{s}
\end{figure}

In Fig. \ref{pe}, we compare the model results for the pressure
and the energy density with the lattice data. Values are
normalized to those in the Stefan-Boltzmann (SB) limit. Both
$\varepsilon/\varepsilon_{\rm SB}$ and $P/P_{\rm SB}$ increase
fast and  monotonically with the temperature above $T_c$, but, as
we see, the Stefan-Boltzmann limit is not saturated up to $3T_c$.
Two sets of lattice data are similar qualitatively, but old
data~\cite{Ka89} are appreciably higher than the new ones
\cite{Pa09} at $T\gsim 1.5T_c$. Note that the new lattice data are
extended to the region of the glueball phase, $T<T_c$. Our QP
model is in a reasonable agreement with the new lattice data,
except for the narrow vicinity to the left of $T_c$, where model
predictions are evidently below the lattice points [see insertions
in Fig. \ref{pe}(a) and Fig. \ref{pe}(b)].
 Due to large glueball masses, this result is not
changed if one adds the next two to three glueball states to our
two lowest-lying glueball states, although there exist statements
claiming that a good agreement with lattice data near $T_c$ can be
reached only if the whole high-lying glueball spectrum of the
Hagedorn-type~\cite{Me09} or glueball condensate~\cite{Bu09} is
additionally included.

The interaction measure or trace anomaly $(\varepsilon-3P)/T^4$
and the reduced enthalpy $(\varepsilon+P)/(\varepsilon_{\rm
SB}+P_{\rm SB})$ are demonstrated in Fig.~\ref{s}(a) and
Fig.~\ref{s}(b), respectively. The presence of a nonzero trace for
the energy-momentum tensor relates to the breaking of the scale
and conformal invariance. Again, a nice overall agreement is
observed between the QP model and the new set of the lattice
data~\cite{Pa09}, including  a region near $T_c$, for $T>T_c$. The
reduced enthalpy in Fig. \ref{s}(b) for the pure gluon system is
just the reduced entropy, $s/s_{\rm SB}$, which is thereby also
reproduced by our QP model. Thus, we see that the developed QP
model successfully describes the thermodynamic properties of the
glue matter.

\section{Calculation of viscosity coefficients}

In principle, it is possible to compute the shear and bulk
viscosities directly from GD at finite temperature using Kubo
formulas. However, in practice, this is quite difficult because GD
is generally a strongly interacting theory with an unknown
mechanism of the confinement. Essential assumptions of our kinetic
approach are that quasiparticles are well defined,  elementary
interactions are local, and the dynamics can be described in the
relaxation-time approximation.

Derivation of viscosity coefficients  starts with the expression
for the energy-momentum tensor for quasi-free \footnote{The
contribution of the QP interaction to the variation of the
energy-momentum  is assumed to be small in the model under
consideration, and therefore it is disregarded.} boson
quasiparticles of spices $a$~:
\begin{eqnarray}\label{Tmunua}
T^{\mu\nu}_a =\int d\Gamma\left\{ \frac{p_a^\mu p_a^\nu}{E_a}F_a
\right\}~,
\end{eqnarray}
\begin{eqnarray}
d\Gamma &=&
 d_a \frac{d^3\vec {p}_a}{(2\pi)^3}
~,\quad  p_a^\mu = (E_a (\vec{p}_a ,\vec{r}),\vec{ p}_a)~,
 \nonumber
\end{eqnarray}
where $d_a$ is the degeneracy factor. The QP distribution function
$F_a$ fulfills the QP kinetic equation. We assume that gluon and
glueball masses are given by Eqs. (\ref{m_g}) and (\ref{Mglb}),
respectively. The QP energy   is determined by
 \be
E_a(\vec{p})=\sqrt{\vec{p}^{\,\,2} +m_a^{2}(T, F_a )}~.
 \label{qe}
\ee

Below, we consider only collisional sources of the viscosity.
Applying the relaxation time approximation to the relativistic QP
kinetic equation, we arrive at the expression for the variation of
the energy-momentum tensor (\ref{Tmunua}) near the local
equilibrium state:
 \be
 \label{varenmom} \delta T^{\mu\nu} =-\sum_a\int
d\Gamma\left\{ \tau_a\frac{p_a^\mu  p_a^\nu}{E_a^2} \
p_a^\kappa\pd_\kappa F_a \right\}_{\rm loc.eq.} ~,
 \ee
 where $\tau_a$ denotes the relaxation time of the given
species, which generally depends on the QP momentum $\vec{p}_a$.
The local equilibrium distribution function for a boson is as
follows:
 \be
 F^{\rm loc.eq.}_a (p_a ,x_a )=\left[e^{p^{\mu}_a u_{\mu}/T}- 1
 \right]^{-1}~,
 \label{leqdf}
 \ee
$u^\mu \simeq (1, \vec{u})$ for $|\vec{u}|\ll 1$. Performing a
variation in (\ref{varenmom}), we did not vary quantities   that
may depend on the distribution function only implicitly, such as
$E_a$, since by doing this  one may arrive at the relaxation-time
form of the QP kinetic equation. Besides, in the gluon-glueball
model used here, only equilibrium values  $m_a (T)$ are known and
we are actually not able to find $\delta E_a [F]$.

The shear and bulk viscosities can be expressed through
the variation of the energy-momentum tensor  as follows:
 \be
 \label{vis}
\delta T_{ij}&=&-\zeta \ \delta_{ij}{\vec{\nabla}}\cdot\vec{u}-\eta
\ W_{ij},\\ \nonumber {\rm with} \quad
W_{kl}&=& \pd_ku_l+\pd_lu_k-\frac23\delta_{kl} \ \pd_iu^i~.
 \ee
Here and below, Latin indices run $1,2,3$. To find the shear
viscosity,  we set $i\neq j$ in (\ref{vis}). To find the bulk
viscosity, we  substitute $i=j$ in (\ref{vis}) and use that
$T^{ii}_{\rm loc.eq}=3P_{\rm loc.eq}$.

Taking derivatives  $\partial F_a^{\rm loc.eq.}/\partial x^{\mu}$
in Eq. (\ref{varenmom})  and using (\ref{vis}) as a definition of
viscosity coefficients, by straightforward calculations we find
expressions (see ~\cite{SR9,SR10,KTV09,KTV10}) for the  shear
viscosity,
 \be\label{shear} \eta=\frac{1 }{15T}\sum_a \limits \int
d\Gamma\,\tau_a \frac{\vec{p}^{\,4}_a}{E^2_a}\, F^{\rm eq}_a \;
\left(1+ F^{\rm eq}_a \right) \ee and for the bulk viscosity
\footnote{We have checked that the Landau-Lifshitz condition
$\delta T^{00}=\sum_a\int d\Gamma \tau_aE_aQ_a=0$ is satisfied in
our QP model.},
\begin{eqnarray}
\label{bulk} \zeta=-\frac{1}{3T}\sum_a \limits \int
d\Gamma\,\tau_a \; \frac{\vec{p}^{\,2}_a}{E_a} \; F^{\rm eq}_a \;
\left(1+ F^{\rm eq}_a\right)Q_a ,
\end{eqnarray}
where the EOS-dependent $Q_a$ factor is given by
\begin{eqnarray}
 \label{Qa} Q_a =
-\left\{\frac{\vec{p}_a^{\,2}}{3E_a}-c_s^2 \left[ E_a
-T\frac{\pd E_a}{\pd T} \right]
\right\}
\end{eqnarray}
and $c_s^2=\frac{\pd P}{\pd \epsilon}$ is the speed of sound squared.

 Simplifying, instead of the momentum-dependent value $\tau_a$, one may
 use the averaged  partial relaxation time
${\tilde\tau}_a$ related to the cross section   as
 \be
  {\tilde \tau}^{-1}_a (T)
=\sum_{a^{'}} n_{a^{'}} (T)\left<v_{aa^{'}}
(v_{aa^{'}}) \right>,
 \label{tau}
 \ee
where $n_{a^{'}}$ is the particle density of $a^{'}$ species,
$\sigma^t_{aa^{'}}=\int d\cos \theta \ d\sigma(aa^{'}\to
aa^{'})/d\cos \theta \ (1-\cos \theta) $ is the transport cross
section, in general,  accounting for in-medium effects, and
$v_{aa^{'}}$ is the relative velocity of two colliding particles
$a$ and $a^{'}$ in the case of binary collisions. Angular brackets
denote a quantum-mechanical statistical average over an
equilibrated system. However, one should bear in mind that
averaged values ${\tilde\tau}_a^{-1}$ given by Eq. (\ref{tau})
yield only a rough estimate for the values ${\tau}^{-1}_a$.

\section{Results for viscosities}

Below, the shear and bulk viscosities are calculated with the help
of Eqs. (\ref{shear}) and (\ref{bulk}), respectively. The only
quantity that should still be specified is the relaxation time
$\tilde{\tau}_a$.

 Calculations of
the relaxation time ${\tilde\tau_a}$ of partons already in the
lowest order in the running coupling constant $g^2$ require
summation of infinitely many diagrams. Resummation of the hard
thermal loops  results in the width ${\tilde\tau}^{-1}$ of partons
$\sim g^2T \ln (1/g)$ \cite{Pi89}. Based on this fact, the
following parametrization was used for gluons~\cite{PC05,Pe01}:
 \be
{\tilde\tau}_g^{-1}=N_c \ \frac{g^2T}{4\pi} \ \ln \frac{2c}{g^2}~,
 \label{tau_g}
 \ee
 with the strong interaction coupling constant (\ref{coupl}) and
a tunning parameter $c$. The relaxation
time for a mixture of scalar and tensor glueballs was estimated
according to Eq. (\ref{tau}) assuming the glueball scattering
cross section $\sigma_{gb}=$30 mb to be isotropic.

\begin{figure}[thb]
\includegraphics[height=5.50truecm,clip]{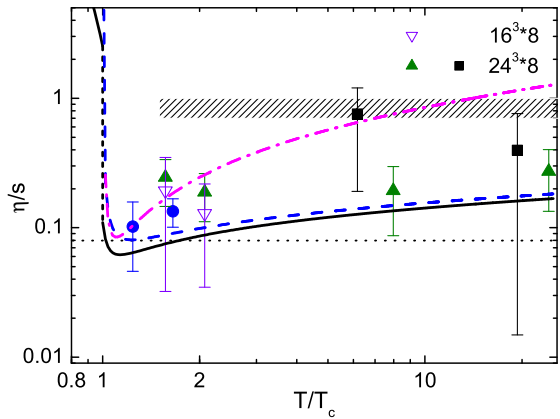}
\caption{The ratio of the shear viscosity to entropy density for a
pure glue matter. Solid  and dashed lines  are for our QP
two-phase gluon-glueball model with two different choices of the
coefficient $c$ in Eq. (\ref{tau_g}); see the text for details.
The vertical short-dashed line joins two boundary points of the
mixed gluon-glueball phase. The dot-dashed line shows viscosity
calculations with the relaxation time $\tilde{\tau}_g$ given by
Eq. (\ref{Kam}).  The horizontal dotted line is the
$\eta/s=1/4\pi$ bound. The lattice gauge $SU(3)$ data with
$16^3\cdot8$ and $24^3\cdot8$ lattice are from Refs.~\cite{SN07}
 (triangles and squares) and \cite{Ma07}
 (filled circles). The shaded region corresponds to the
perturbative result (cited from~\cite{NS05}).
 }
 \label{etas}
\end{figure}

Comparison between the GD lattice data \cite{SN07,Ma07} and our QP
results for the shear viscosity of the glue matter is presented in
Fig. \ref{etas}. The magnitude of the $\eta/s$ ratio in our model
is defined mainly by the value of the relaxation time
(\ref{tau_g}).  The solid line shows results of our calculation
provided we use the recommended value $c=$14.4, though our
parameters of Eq. (\ref{coupl}) are slightly different from those
used in Ref.~\cite{PC05}. This $c$-value was tuned in
Ref.~\cite{PC05} to the old lattice data for thermodynamic
quantities~\cite{BEG95}.  As we see, the $\eta/s$ ratio gets
discontinuity at $T=T_c$ with more than by an order of magnitude
lower value at $T\to T_c+0$ (in gluon phase) than at $T\to T_c-0$
(in the glueball phase). Also, the solid curve lies  reasonably
close to the points (filled circles, triangles, and squares) and
its value in the minimum is slightly below the AdS/CFT (Anti de
Sitter/Conformal Field Theory) $1/4\pi$ bound ~\cite{KSS03}
(compare with the dotted curve). Preserving the form of the
relaxation time (\ref{tau_g}), we can still increase $\eta/s$
values by tuning the parameter $c$. Taking $c = 11.44$, we achieve
the limit case $\tilde{\tau_g}^{-1}\to $ 0 for $T\to T_c+0$ (full
transparency). In this case (see the dashed line in Fig.
\ref{etas} for $T>T_c$),
 we may reach  a slightly better overall agreement
 with the lattice data~\cite{SN07,Ma07},
 and the  $1/4\pi$ bound is achieved at the minimum. Varying the
 $c$-value in the interval  $11.44<c<14.4$,
 one may simulate different values of the $\eta/s$ jump at $T=T_c$,
but for temperatures $T\gsim 1.5 T_c$ the $\eta/s$
 ratio changes only slightly, demonstrating  a slow
increase with the growing temperature. Thus, bearing in mind large
error bars in the  lattice data, we are able to conclude that the
results of the two-phase gluon-glueball model developed here  are
consistent with the existing lattice results~\cite{SN07,Ma07}. The
perturbative regime (see the shaded region) is not achieved up to
very high temperatures.

The $\eta/s$ ratio for the pure gluon phase in the range of
 $T\sim (1-2) T_c$ was
also evaluated in Ref. \cite{BKR09}. The model employs the QP {\it
Ansatz}  for EOS successfully tested to describe old lattice
results~\cite{NS05}. In Ref. \cite{BKR09}, viscosity is treated by
means of the kinetic theory for gluon quasiparticles. It is of
interest that the model,  being consistent with the old (and less
accurate) lattice data for viscosity~\cite{NS05} and
thermodynamics~\cite{Ka89} that overestimate pressure at $T\gsim
1.5T_c$
 (as follows from the comparison with new data, see Fig.~\ref{pe}),
predicts a stronger temperature dependence of $\eta/s$ at $T>T_c$
than our model, which in turn is consistent with the new lattice
data \cite{SN07,Ma07}. The crucial point here is  that  the gluon
relaxation time is defined essentially
differently:
 \be \label{Kam}
{\tilde{\tau}}^{-1}_{\rm BKR} =a_\eta /(32\pi^2)T \
g^4\ \log(a_\eta \pi / g^2)~,
 \ee
where $a_\eta=$6.8. Here $ {\tilde{\tau}}^{-1}_{\rm BKR}\propto
g^4$, coming from the $\alpha_s^2\sim g^4$ dependence of the
gluon-gluon transport cross section $\sigma^t_{aa^{'}}$ in the
relaxation-time expression (\ref{tau}),  as   was estimated in the
early work of Hosoya and Kaiantie~\cite{HK}, whereas the above
used ${\tilde{\tau}}^{-1}_g\propto g^2$~\cite{PC05,Pe01}.

The $\eta/s$ ratio obtained with  the relaxation time (\ref{Kam})
is  plotted  in Fig. \ref{etas} by the dot-dashed line.  This
result for $T\lsim 2T_c$ recovers that of Ref.~\cite{BKR09}, but
it differs significantly from those calculated with Eq.
(\ref{tau_g}). Using recent lattice results for higher
$T$~\cite{SN07}, let us try to make a choice between two
parametrizations of relaxation times (\ref{tau_g}) and
(\ref{Kam}). We observe that for $T\gsim 10T_c$, shear viscosity
calculations with (\ref{Kam}) demonstrate a noticeable growth
exceeding lattice data and even a perturbative estimate
$(\eta/s)_{\rm pert}\approx 0.8-1.0$ for $T\simeq (2 - 20) T_c$.
The perturbative result is taken from Fig. 2 of \cite{NS05}
($\eta$ was calculated in \cite{AMY} and $s$ in \cite{BIR}). In
contrast, predictions of our QP model  with relaxation time
(\ref{tau_g}) are in reasonable agreement with the lattice results
and do not contradict perturbative estimates. Recently, the
relaxation time $\tilde{\tau}_g$ was  estimated in
Ref.~\cite{XG08} according to Eq. (\ref{tau})
 from an analysis of  cross sections of the $gg\to
gg$ and $gg\to ggg$ processes. It was found that $\eta/s$=0.13 and
0.076 for values $\alpha_s=$0.3 and 0.6, respectively [which
correspond to temperatures $T/T_c=$2.6 and 1.36 provided Eq.
(\ref{coupl}) is used]. If these points were plotted in
Fig.~\ref{etas}, they would be quite consistent with our QP model
results. This can be considered as an additional numerical
argument in favor of using Eq.  (\ref{tau_g}) as a rather
appropriate phenomenological expression. However, we point out
that the above arguments do not allow us to choose between
functional dependencies (\ref{tau_g}) and (\ref{Kam}). One could
also use expression (\ref{Kam}) with a smaller coefficient
$a_{\eta}$, choosing the latter value to fit the lattice results
for the $\eta/s$ ratio. Since uncertainties in the modern lattice
data are large, we will not do such an additional tuning
restricted by the results that we have demonstrated in Fig.
\ref{etas}.

\begin{figure}[thb]
\includegraphics[height=5.5truecm,clip]{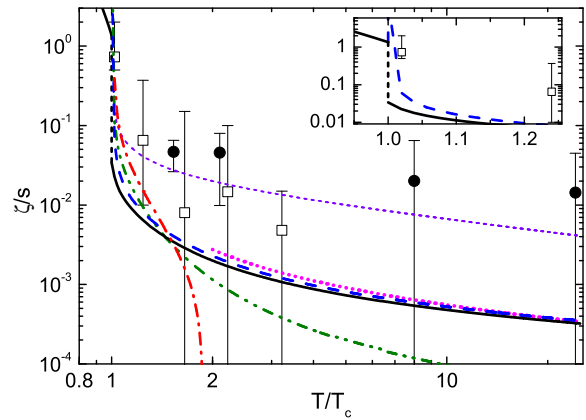}
\caption{ The bulk viscosity to entropy density ratio for a glue
matter. Solid and dashed lines are the results of our QP two-phase
gluon-glueball model with two relaxation times as in Fig.
\ref{etas}. (the vertical short-dashed line joins boundary points
of the mixed phase). The dash-double-dotted line is the
calculation result with Eqs. (\ref{bl2}) and (\ref{tau_g}), and
the dot-dashed one is calculated according to Eqs. (\ref{bl3}) and
(\ref{Kam}). The perturbative estimate (\ref{pert}) is plotted by
the dotted line. Experimental points are from~\cite{Me08} (empty
squares) and \cite{SN07} (filled circles). The thin short-dashed
curve corresponds to a simple approximating dependence
$\zeta/s=0.02/\sqrt{T/T_c-1}$ to guide the eye. }
 \label{Fbulk}
\end{figure}

The measured lattice points for the ratio of the bulk viscosity to
the entropy density are plotted in Fig. \ref{Fbulk} together with
different model results. In a broad range of temperatures, the
global behavior of lattice data can be roughly approximated as
$\zeta/s =0.02/\sqrt{T/T_c-1}$ (see the short-dashed curve).  The
reduced bulk viscosity $\zeta/s$ calculated  in our two-phase
gluon-glueball model following Eqs. (\ref{bulk}) and (\ref{tau_g})
is shown by the solid line for $c=14.4$ and by the dashed line for
${\tilde{\tau}}^{-1}_g$ vanishing at $T_c$. Values of $\zeta/s$
for both curves noticeably underestimate the corresponding values
on the approximating short-dashed curve, typically by an order of
magnitude. Nevertheless, the  shape of the curves is similar to
that given by the approximating  curve. Singularity at $T\to
T_c+0$ demonstrated by the  dashed line (see insertion in Fig.
\ref{Fbulk}) is due to the divergence of $\tilde\tau_a$ in this
limiting case.

The bulk viscosity (\ref{bulk}) includes a rather complicated
factor $Q_a$ depending on the EOS used. Using the energy
conservation for a system with temperature-independent masses of
particles,  one may present the result (\ref{bulk}) as follows:
 \be \label{bl1}
  \zeta=\sum_a\frac{d_a}{T}\int\frac{d^3p}{(2\pi)^3}\tau_aF^{\rm eq}_a(1+
  F^{\rm eq}_a) \left[\frac{\vec p^{\,2}}{3E_a}-c_s^2E_a\right]^2.
 \ee
 For a single-component gas, this expression coincides exactly with the
 25-years-old result of Gavin~\cite{G85}.

Chakraborty and Kapusta (CK)~\cite{CK10} presented another
expression \footnote{Here we denote the relaxation time
$\bar{\tau}_a$ to emphasize its difference from the quantity
$\tau_a$ used in Eq. (\ref{bulk}); see discussion in the Appendix.
},
 \be \label{bl2}
  \zeta_{\rm CK}=\sum_a\frac{d_a}{T}\int\frac{d^3p}{(2\pi)^3}
  {\bar{\tau}_a}F^{\rm eq}_a(1+ F^{\rm eq}_a)Q_a^2
 \ee
which differs from (\ref{bulk}) but also reduces to  (\ref{bl1})
for $m_a=const$. Note that they also disregard the QP interaction
term in the energy-momentum tensor; see Eq. (99) of their work.
The reasons for the differences between (\ref{bulk}) and
(\ref{bl2}) are discussed in the Appendix. The dash-double-dotted
line in Fig. \ref{Fbulk} demonstrates the $\zeta_{\rm ChK}/s$
ratio following
 Eq. (\ref{bl2}) with the  relaxation time $\bar{\tau}_g =
 \tilde{\tau}_g$ given by Eq.  (\ref{tau_g})
 (for $\tilde \tau^{-1}_g \to 0$ at $T\to T_c+0$).
 We see that  Eq. (\ref{bl2})
  yields a strong $T$ suppression of the bulk viscosity at
$T\gsim 1.5T_c$, as compared to that given by Eq. (\ref{bulk})
 (compare the
   dash-double-dotted and dashed   lines in  Fig. \ref{Fbulk}).

A somewhat different expression for $\zeta$  was used by Bluhm,
K\"ampfer, and Redlich (BKR)~\cite{BKR09}. In their  model, the
bag constant $B$  is a  functional of the non-equilibrium
distribution function. They found
 \be \label{bl3}
  \zeta_{\rm BKR}&=&\sum_a\frac{d_a}{3T}\int\frac{d^3p}{(2\pi)^3}
  \frac{\tau_a}{E_a}F^{\rm eq}_a(1+ F^{\rm eq}_a)
  \nonumber\\  &\times& Q_a \left[
  m_a^{2}(T)-T\frac{dm_a^{2}(T)}{dT}\right]~.
 \ee
 Here it was assumed that the QP interaction
 contributes to the energy-momentum tensor.
 Thereby, compared to (\ref{bulk}) there appeared  the second term
 $T{dm^2_a(T)}/{dT}$ in the square brackets of Eq.~(\ref{bl3}).
For constant masses, the latter equation is also reduced to
(\ref{bl1}). Numerical calculations  with (\ref{bl3}) (see the
dot-dashed line in Fig. \ref{Fbulk}) give rise  to the $\zeta_{\rm
BKR}/s$ ratio, which dramatically decreases, being in large
discrepancy with both the above models and the lattice data for
$T>1.5 T_c$. For $T>1.9T_c$, Eq.~(\ref{bl3}) becomes invalid,
providing negative values.

A perturbative estimate~\cite{Me08} gives
 \be \label{pert}
(\zeta/s)_{\rm pert} \approx 0.02\alpha_s^2
 \ee
 for $0.06\le\alpha_s\le 0.3$. Applying the $T$-dependent
 coupling constant
(\ref{coupl}) for $\alpha_s=g^2(T)/4\pi$, we get a perturbative
estimate of the bulk viscosity to entropy density ratio (plotted
by the dotted line in Fig. \ref{Fbulk}).  As is seen, in the
region of its applicability,  Eq. (\ref{pert}) produces   only
slightly larger values of $\zeta/s$ than those given by our QP
model. The new  lattice GD calculations demonstrate a significant
increase of the ratio $\zeta/s$ at  approaching the critical point
 ($\zeta/s \simeq 0.5\div 2$ at $T=1.02~T_c$). These values are
 reproduced  neither by our QP
model using relaxation time Eq. (\ref{tau_g}) nor by the
approximating short-dashed curve exploiting a simple $T$
dependence of the $\zeta/s$ ratio.

 Comparing results presented in Figs. \ref{etas} and \ref{Fbulk}, we see that  in the gluon phase in a narrow vicinity
of the critical point [for $(T-T_c)/T_c \lsim 0.1$],   the ratio
$\zeta/\eta \gsim 0.1$ reaches the value $\zeta/\eta \simeq 0.3$
for $T\to T_c+0$. The ratio sharply decreases with an increase of
the temperature up to values
  $\zeta/\eta\sim 10^{-2}\div 10^{-3}$ for $T> 2T_c$. The smallness of this
ratio controls the violation of the conformal symmetry.

\section{Conclusions}

A quasiparticle approach has been applied to the  $SU(3)$ glue
matter with temperature-dependent masses.  Matching the pure gluon
and glueball phase descriptions by means of the Gibbs conditions
allows one to describe successfully this system in a
thermodynamically consistent way both above and below the critical
temperature $T_c$. For thermodynamic characteristics,  the
quasiparticle model results  are in good agreement with the latest
lattice data.

The constructed equation of state was used to calculate the
shear and
 bulk viscosities in the relaxation-time approximation in a wide
 temperature range.   The magnitudes of the shear and bulk
viscosities are mainly determined by the  value of the relaxation
time.
We exploited two different values for the relaxation time that
have been used in the literature; see Eqs. (\ref{tau_g}) and
(\ref{Kam}). With the relaxation time (\ref{tau_g}), the shear
viscosity to entropy density ratio $\eta/s$ fits rather well the
scant lattice data. We found that the ratio $\eta/s$ undergoes a
discontinuity at the critical temperature $T=T_c$. At $T$ slightly
above $T_c$ the ratio $\eta/s$ has a minimum, the value of which
is close to the AdS/CFT bound $1/4\pi$. Then $\eta/s$ increases
with the subsequent rise of the temperature. The bulk viscosity to
entropy density ratio $\zeta/s$  also has
 a break at $T_c$. Then it monotonically decreases with the
temperature increase.
Although the calculated $\zeta/s$ ratio essentially underestimates
the upper limits given by the corresponding lattice data, its
temperature dependence  is well described.

Within our model, the ratio $\zeta/\eta \simeq 0.3 $ at $T\to
T_c+0$ and it sharply decreases with  the rising temperature up to
values $\zeta/\eta \sim 10^{-2}\div 10^{-3}$ for $T>2T_c$.

 We point out that although our QP
model describes well the thermodynamical characteristics
calculated on the lattice it does not guarantee that all important
physics is incorporated, especially in a description of the
vicinity of the critical point, where fluctuation effects are
increased. Using  the averaged value for the relaxation time can
be considered only as a very rough approximation. The dependence
of the relaxation time on the coupling constant $g$ is also not
well defined. Simplifying, all existing approaches to the
evaluation of the bulk viscosity use some {\it Ans\"{a}tze}
reductions yielding different results. We used the kinetic
approach with the collisional source of the viscosity
 disregarding other possible sources~\cite{PP06}.
One such source is associated with the presence of a soft mode
\cite{ML37} in the vicinity of the second and weak first-order
phase-transition critical points.
Also, since statistical error bars are very large, new more
certain lattice data are required to draw a more definite
conclusion on the agreement or disagreement of the calculated
$\zeta/s$ ratio with the lattice results.

  \vspace{3mm} \begin{center} {\bf Acknowledgements}
\end{center}
We are thankful to D.~Blaschke, W.~Cassing, M.~Gorenstein,
Yu.~Ivanov, E.~Kolomeitsev, N.~Koshelev and V.~Skokov   for useful
discussions. This work was supported in part by the DFG grant WA
431/8-1, RFFI grants 08-02-01003-a, Ukrainian-RFFI grant
09-02-90423-ukr-f-a, and a Heisenberg-Landau grant.

\section{Appendix}

 Deriving kinetic coefficients, the
 authors of \cite{SR9,SR10,KTV09,KTV10}
used the relaxation-time approximation to the kinetic equation
presenting the collision integral as
 \be\label{col}
St F_a =-\delta F_a/\tau_a [F^{\rm loc.eq.}],\ee
 where
 \be\label{delF} \delta F_a =F_a (E_a [F])
-F_a^{\rm loc.eq.}(E_a [F^{\rm loc.eq.}]),\ee
 see, e.g., Eqs.
(38) and (40) in \cite{KTV10}. Here it was assumed that the
collision term should be zero for the global and  local
equilibrium states, i.e., for $F_a =F_a^{\rm loc.eq.}(E_a[F^{\rm
loc.eq.}])$.

Then after setting $F_a =F_a^{\rm loc.eq.}(E_a[F^{\rm loc.eq.}])$
on the left-hand side of the kinetic equation, one  finds
 \be\label{dF}\delta F_a
=-\frac{\tau_a [F^{\rm loc.eq.}]}{E_a [F^{\rm loc.eq.}]
}p_a^{\mu}\frac{\partial F_a^{\rm loc.eq.}(E_a [F^{\rm
loc.eq.}])}{\partial x_a^{\mu}},\ee
 see Eq. (2.3) in \cite{SR9} and Eq. (42) in \cite{KTV10}. We stress
that all quantities on the right-hand side of this equation
including the relaxation time $\tau$ are expressed in terms of the
{\em local equilibrium distribution functions}.

To derive expressions for the shear and bulk viscosities
(\ref{shear}) and (\ref{bulk}), one presents spatial components of
the variation of the energy-momentum tensor of quasiparticles
(\ref{Tmunua}) as
 \be\label{Redlich}
 &&\delta T^{ik}=\sum_{a} \int d\Gamma \frac{p_a^i p_a^k}{E_a
[F]}\delta {F}_a -\sum_{a} \int d\Gamma \frac{p_a^i p_a^k
{F}_a^{\rm loc.eq.}}{E_a^2 [F^{\rm loc.eq.}]}\delta E_a\nonumber\\
 &&\rightarrow \sum_{a} \int d\Gamma \frac{p_a^i p_a^k}{E_a
[F^{\rm loc.eq.}]}\delta {F}_a .\ee
 To avoid cumbersome expressions we omitted antiparticle terms.
The reduction done in the second line in Eq. (\ref{Redlich}) is
actually an {\it Ansatz}: we vary only
 the distribution function and do not vary quantities that
depend on the distribution function  implicitly (through
phase-space integrals incorporating the distribution function),
i.e.,  $\delta E_a [ \delta F]$ are set zero. This reduction is in
the spirit of the relaxation-time approximation to the kinetic
equation, where the momentum-dependent relaxation-time parameter
is replaced in actual calculations by an averaged value. Note that
dropping the $\delta E$ term, we actually ignore a sub-leading
term in the case of a weak coupling constant and/or for a very
dilute system, see Eq. (5.22) of \cite{Jeon}. The distribution
function $\delta F_a$ counted from the local equilibrium value
enters the expression for $\delta T^{ik}$, the expression for
$\delta T^{00}=0$ [see (45) and (50) in \cite{KTV10}], and
expressions for the viscosities. Thus one can easily compute
kinetic coefficients knowing thermodynamic quantities in the local
rest frame $\vec{u}=0$.

However, we should note that  in the  QP  Fermi liquid theory
following the work of Abrikosov and Khalatnikov \cite{AKh}, one
usually uses a different procedure to obtain transport
coefficients; see  \cite{BaymPethick} for details. One exploits
that in the original Landau collision term, the following
combination enters:
 \be \label{delta}
&&\delta \left(\sum_{a} E_a [F]\right)\left\{ F_1 (E_1) F_2
(E_2)(1\mp F_3 (E_3))(1\mp F_4 (E_4))\right.\nonumber
\\&&\left. -F_3 (E_3) F_4 (E_4) (1\mp F_1(E_1))(1\mp F_2 (E_2))\right\},
 \ee
 where $E_a$ are functionals of {\em the exact
non-equilibrium distribution function}, $a=1,2,3,4$. The term in
curly brackets
 is zero not only for $F_a
=F_a^{\rm loc.eq.}[E_a(F^{\rm loc.eq.})]$ but also for $F_a^{\rm
loc.eq.} (E_a [F])$. Thereby, $St F_a^{\rm loc.eq.}(E_a [F]) =0$.
 Thus introducing
 \be\label{delFtilde}\delta \widetilde{F}_a =F_a
(E_a [F])-F_a^{\rm loc.eq.} (E_a [F])\ee
 in the relaxation-time
approximation, we may rewrite the collision term as
 \be\label{stF} St F_a
=-\delta \widetilde{F}_a/\bar{\tau}_a [E(F)].\ee

The quantity $\bar{\tau}[F]$ entering Eqs.  (\ref{stF}) and
(\ref{bl2}) depends on the unknown exact non-equilibrium
distribution function, since  the $\delta$-function term in the
collision integral and the local equilibrium distributions there
continue to depend on exact energies in this approach. If we want
to calculate the value of $\tau [F^{\rm loc.eq.}]$ entering Eqs.
(\ref{shear}) and (\ref{bulk}) using Eq. (\ref{stF}), we should
still expand $E[F]$ in (\ref{stF}) near the known value $E[F^{\rm
loc.eq.}]$ everywhere including the $\delta$-function term in the
collision integral.

From the left-hand side of the kinetic equation, one gets
 \be\label{dFtilde}\delta \widetilde{F}_a
=-\frac{\bar{\tau}_a [F]}{E_a [F] }p_a^{\mu}\frac{\partial
F_a^{\rm loc.eq.}(E_a [F])}{\partial x_a^{\mu}}.
 \ee
Then one may use a simple expression for $\delta T^{ik}$,
 \be\label{Abr}\delta T^{ik}=\sum_{a} \int
d\Gamma \frac{p_a^i p_a^k}{E_a [F]}\delta \widetilde{F}_a
 \ee
since now variations are always performed at fixed $E_a$. As
above, the QP interaction term is omitted. Comparing the second
line of Eqs. (\ref{Redlich}) and (\ref{Abr}), we see that
disregarding the implicit dependence $E[\delta F]$, Refs.
\cite{SR9,SR10,KTV09,KTV10} actually do not distinguish between
distributions $\delta F$ and $\delta \widetilde{F}$.

 Then in both considered approaches, one
uses the exact relation $E_a [F] =\delta T^{00}/\delta F_a ,$
i.e., that the variation of the energy  is determined through
$\delta F_a$ as
 \be\delta T^{00}&=&\sum_{a}\int d\Gamma E_a [F]\delta F_a \nonumber\\
 &=& \sum_{a}\int d\Gamma E_a [F^{\rm loc.eq.}]
 \delta F_a +O[(\delta F)^2]
.
 \ee
Following (\ref{delF}) and (\ref{delFtilde}), we have
 \be\label{FFeq} \delta F -\delta \widetilde{F}=\frac{\partial F^{\rm
 loc.eq.}(E[F^{\rm loc.eq.}])}{\partial E}\delta E .
 \ee
Furthermore, instead of using a complicated implicit dependence
$\delta E [\delta F]$ with $\delta F$ given by Eq. (\ref{dF}),
which would be the fully correct procedure, Ref. \cite{CK10} uses
the {\it Ansatz} relations  [see Eq.  (102) of that work]
 \be\label{FF}
&&\delta \widetilde{F}=\mbox{exp}\left\{-\frac{E[F^{\rm
loc.eq.}]}{T[F^{\rm loc.eq.}]}\right\}  \frac{E^{\rm
loc.eq.}}{T_{\rm loc.eq.}^2} \delta T ,\\ &&\delta E= \frac{\delta
T}{E}\frac{mdm}{dT}=\frac{\delta
\widetilde{F}}{F}\frac{T^2}{E^2}\frac{ mdm}{dT},\nonumber
 \ee
 which
assume that the distribution function in a nonequilibrium state
has the form $F=e^{-E[F]/T[F]}$ in the Boltzmann limit $F\ll 1$.
Thus, although Ref. \cite{CK10} distinguishes between
distributions $\delta F$ and $\delta \widetilde{F}$, it uses  very
special relations (\ref{FF}), which might be incompatible with
$\delta E(m[\delta \widetilde{F}])$ as it follows from    Eqs.
(\ref{dFtilde}) and (\ref{qe}).

To find bulk viscosity, one further expresses $F_a =F_a^{\rm
loc.eq.}(E_a [F])(1-A_a
\partial_\rho u^{\rho})$, see Ref. \cite{CK10}, and  one
observes that  the shift of the solution
 $ A(E)\rightarrow A(E)-bE$
 generates new solutions of  the
Landau kinetic equation for arbitrary constant $b$. Then one
chooses   $b$
  to explicitly fulfill the Landau-Lifshitz condition
$u_\mu \delta T^{\mu\nu}=0$. Note that this modification of the
solution is quite unnecessary provided  one guarantees that  the
condition $\delta T^{00}=0$ holds in the local rest frame. We have
checked that this condition is satisfied in our QP model.

Finally, within this approach  one arrives at the
 expression (\ref{bl2})   for the bulk
viscosity, which is explicitly positive definite, whereas positive
definiteness of Eq. (\ref{bulk}) is not seen explicitly. However,
we stress once more that {\em all quantities in (\ref{bl2}) still
depend on exact energies, while the way in which the quantities in
(\ref{bulk}) depend on unknown exact distribution functions is
hidden.}
 Thus explicit positive definiteness of expression  (\ref{bl2})
 for $\zeta$ presents actually  only an
apparent improvement. In any case, to use Eq. (\ref{bl2}) in
practical calculations, where only equilibrium quantities are
known, one should replace $E[F]$ by $E[F^{\rm loc.eq.}]$.

 Moreover, we should stress that the
values of the relaxation time in (\ref{bulk}) and (\ref{bl2}) are
different. Since we do not perform complicated microscopic
calculations of the relaxation time but only estimate its average
value, we actually cannot determine whether (\ref{bulk}) or
(\ref{bl2}) is more preferable, and we may use both of them.

Note that Eq. (\ref{bl3})  is derived for a different model, where
the QP interaction
 contributes to the energy-momentum tensor. Also, authors of \cite{BKR09} use
 different value for the relaxation time.

Thus different {\it Ans\"{a}tze} used  in derivation of Eqs.
(\ref{bulk}), (\ref{bl2}), and  (\ref{bl3})  lead to different
values of the bulk viscosity, as is shown in Fig. \ref{Fbulk}.

\end{document}